\documentclass[aps,prl,twocolumn,showpacs,superscriptaddress,groupedaddress,nofootinbib]{revtex4}  
\usepackage{graphicx}  
\usepackage{dcolumn}   
\usepackage{bm}        
\usepackage{amsmath,amssymb}   
\usepackage{aas_macros}
\usepackage{color}
\usepackage{times}
\usepackage{url}
\usepackage{hyperref}
\usepackage{multirow}
\newcommand{\Mpc}{\ensuremath{\,{\rm Mpc}}}

\begin{document}
\widetext

\title{Measurement of Neutrino Masses from Relative Velocities}
\author{Hong-Ming Zhu}
\affiliation{Key Laboratory for Computational Astrophysics,
National Astronomical Observatories, Chinese Academy of Sciences,
20A Datun Road, Beijing 100012, China}

\author{Ue-Li Pen}
\affiliation{Canadian Institute for Theoretical Astrophysics, 60 St. George Street, Toronto, Ontario M5S 3H8, Canada}
\affiliation{Canadian Institute for Advanced Research, CIFAR Program in Gravitation and Cosmology, Toronto, Ontario M5G 1Z8, Canada}

\author{Xuelei Chen}
\affiliation{Key Laboratory for Computational Astrophysics,
National Astronomical Observatories, Chinese Academy of Sciences,
20A Datun Road, Beijing 100012, China}
\affiliation{Center of High Energy Physics, Peking University, Beijing 100871, China}

\author{Derek Inman}
\affiliation{Canadian Institute for Theoretical Astrophysics, 60 St. George Street, Toronto, Ontario M5S 3H8, Canada}

\author{Yu Yu}
\affiliation{Key laboratory for research in galaxies and cosmology,
Shanghai Astronomical Observatory, Chinese Academy of Sciences,
80 Nandan Road, Shanghai 200030, China}
\date{\today}

\begin{abstract}
We present a new technique to measure neutrino masses using their flow
field relative to dark matter.  Present day streaming motions of
neutrinos relative to dark matter and baryons are several hundred
km/s, comparable with their thermal velocity dispersion.  This
results in a unique dipole anisotropic distortion of the matter-neutrino
cross power spectrum, which is observable through the dipole distortion
in the cross correlation of different galaxy populations.  Such a dipole
vanishes if not for this relative velocity and so it is a clean signature for 
neutrino mass. We estimate the size of this effect and find that current and future
galaxy surveys may be sensitive to these signature distortions.
\end{abstract}

\maketitle

\textit{Introduction.}---Neutrinos are now established to be massive,
and the mass differences have been measured, but the mass hierarchy 
and absolute mass values remain unknown \cite{Cahn:2013}.
Precision large scale structure data can be used to measure or constrain
the sum of neutrino masses, as cosmic neutrinos 
with finite masses slightly suppress the growth of
structure on scales below the neutrino thermal free-streaming
scale \cite{Bond:1980,Hu:1997,Saito:2008,Abazajian:2011}.
But the challenge of this method is to conclusively
disentangle the complex and poorly understood baryonic effects as many processes can lead to power suppression on small scales. 
In this Letter, we present an astrophysical effect 
which provides a new way to measure the neutrino masses 
by using a distinct signature in current or future galaxy surveys.

We consider the relative velocity between cold dark matter 
(CDM) and neutrinos. Neutrinos decoupled early in the history of 
the Universe when they were still relativistic, 
but their energy gradually decreased as the Universe
expanded until they behaved as nonrelativistic particles.  At this point
they can cluster under the action of gravity. 
Nevertheless, due to their low masses
the neutrinos can travel relatively large distances (even at low redshifts), and be
perturbed by the underlying gravitational potential 
along their trajectories. The large scale structures can induce 
a significant bulk relative velocity field between CDM and
neutrinos, with typical velocities comparable to the neutrino 
thermal velocity dispersion. As we shall show below,
such a bulk relative velocity field will cause a local 
dipole asymmetry in the CDM-neutrino cross-correlation function.
The concept of dipole asymmetry in correlation functions
was discussed in Ref. \cite{Bonvin:2013} recently. The 
CDM-neutrino cross correlation may be inferred
from the cross-correlation of different galaxy populations, and such a
dipole asymmetry provides a distinctive and robust 
signature of neutrino mass, since such dipole anisotropy would be absent
if not for this effect. 

In this Letter, we delineate the principle of this 
method, make an analytical estimate of the size of this effect,
and then forecast the detectability of this effect in 
a simplified galaxy bias model. 

\textit{The relative velocity.}---We treat CDM and neutrinos as two 
fluids \cite{Shoji:2010} interacting with each other through gravity.  
The CDM particles and neutrinos are collisionless,
nevertheless much of their behavior in gravitational fields can still
be modeled with the introduction of an ``effective pressure,'' 
which takes into account the velocity dispersion or thermal motion 
of the particles \cite{Shoji:2010}. In the fluid approximation, 
the effect of the thermal motion is included in this effective pressure
and only the bulk motion is considered. 
The two fluids have different effective pressure, so they
acquire different densities and velocities even though
they are under the action of the same gravitational field.
We use the moving background perturbation theory 
(MBPT) \cite{Tse:2010} to calculate analytically the evolution of 
the density perturbations and velocities of the two fluids, 
the details of this calculation are given in the Supplemental Material \cite{SuppMat}. 
The basic idea
is to assume that within a certain volume of radius $R$, 
each fluid has a coherent bulk velocity, which 
can be expanded around a background
velocity as $\bm{v}_i(\bm{x},t)=\bm{v}_i^{(bg)}(t)+\bm{u}_i(\bm{x},t)$, where 
$i$ refers to neutrino $(\nu)$ or cold dark matter $(c)$. 
The background velocity $\bm{v}_i^{(bg)}$ is a slowly varying 
velocity long mode. Linear perturbative
calculation then can be applied within the region to obtain the 
cross-correlation of the two fluids.

\begin{figure}[tbp]
\begin{center}
\includegraphics[width=0.48\textwidth]{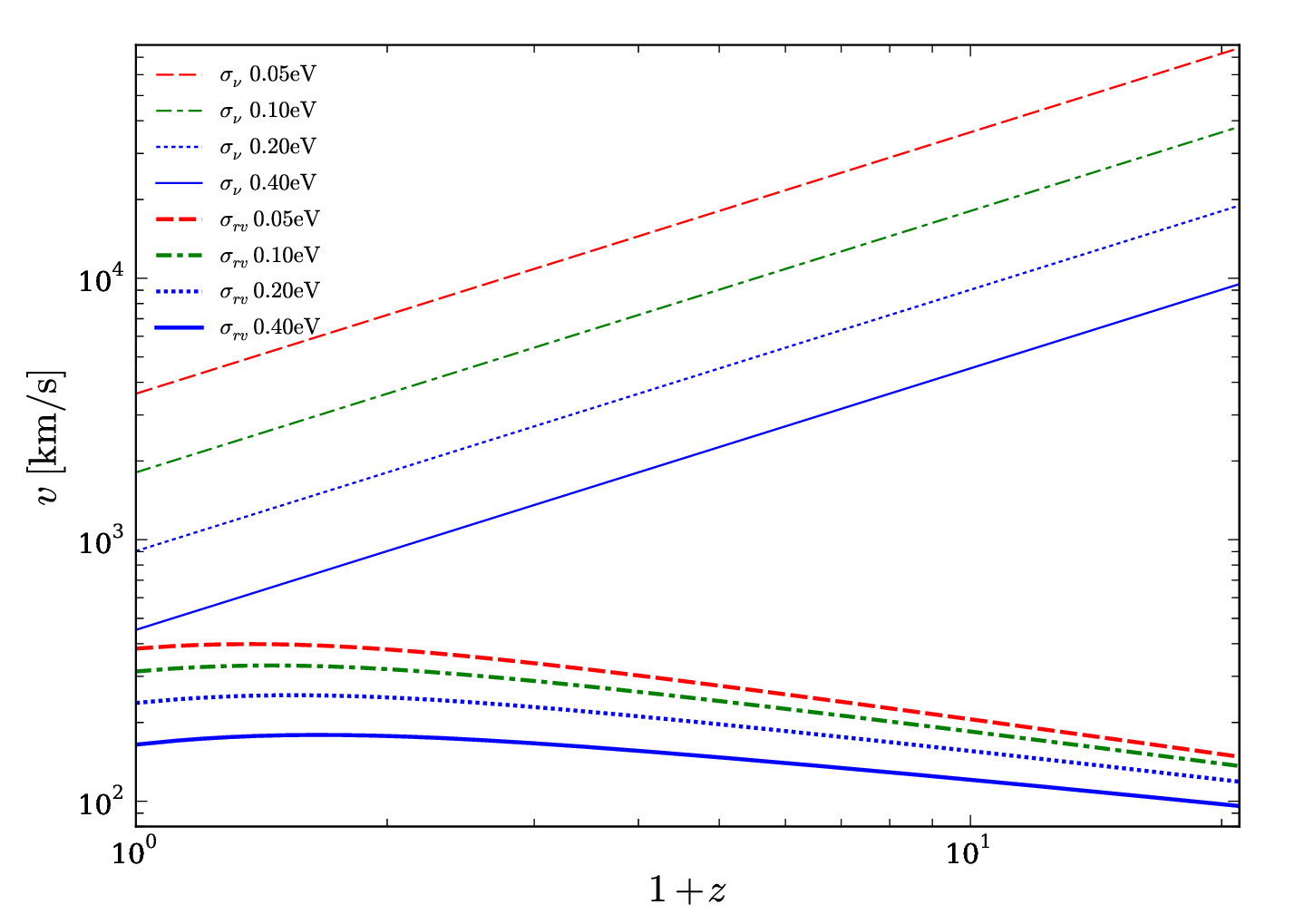}
\end{center}
\vspace{-0.7cm}
\caption{\label{fig:vel} Redshift evolution of the neutrino velocity dispersion 
(the thin lines on top)  and the 
neutrino-CDM relative velocity (thick lines at the bottom) 
for different neutrino masses.
}
\end{figure}

Starting at a high redshift (we use $z=15$ in our calculation) 
when the relative 
bulk mach number is small, we evolve the MBPT equations down to
lower redshifts, and 
obtain the relative velocity field $\bm{v}_{\nu c}(\bm{x}, z)$.
We estimate the variance of this relative velocity analytically 
by taking the ensemble average for the given distribution of primordial
fluctuations:
\begin{equation}
\langle v^2_{\nu c}(z) \rangle
= \int \frac{dk}{k}\Delta^2_{\zeta}(k)\bigg[ \frac{\theta_\nu (k, z) - \theta_c (k, z)}{k} \bigg]^2.
\end{equation} 
where $\Delta^2_\zeta$ is the primordial curvature perturbation spectrum, and
$\theta\equiv\nabla\cdot{\bm v}$ is the velocity divergence. 
We plot the evolution of $\sqrt{\langle v^2_{\nu c}\rangle}$ ($\sigma_{rv}$)
and the neutrino thermal velocity dispersion $\sigma_\nu$
for four neutrino masses in Fig.~\ref{fig:vel}. 
The thermal velocity dispersion of the neutrinos decreases as the 
Universe expands. On the other hand, the bulk
relative velocity as represented by $\sqrt{\langle v^2_{\nu c}\rangle}$ 
grows to its maximum 
at $0<z<1$, then begins to decay. At low redshifts it is 
comparable with the thermal velocity dispersion.

\begin{figure}[tbp]
\begin{center}
\includegraphics[width=0.48\textwidth]{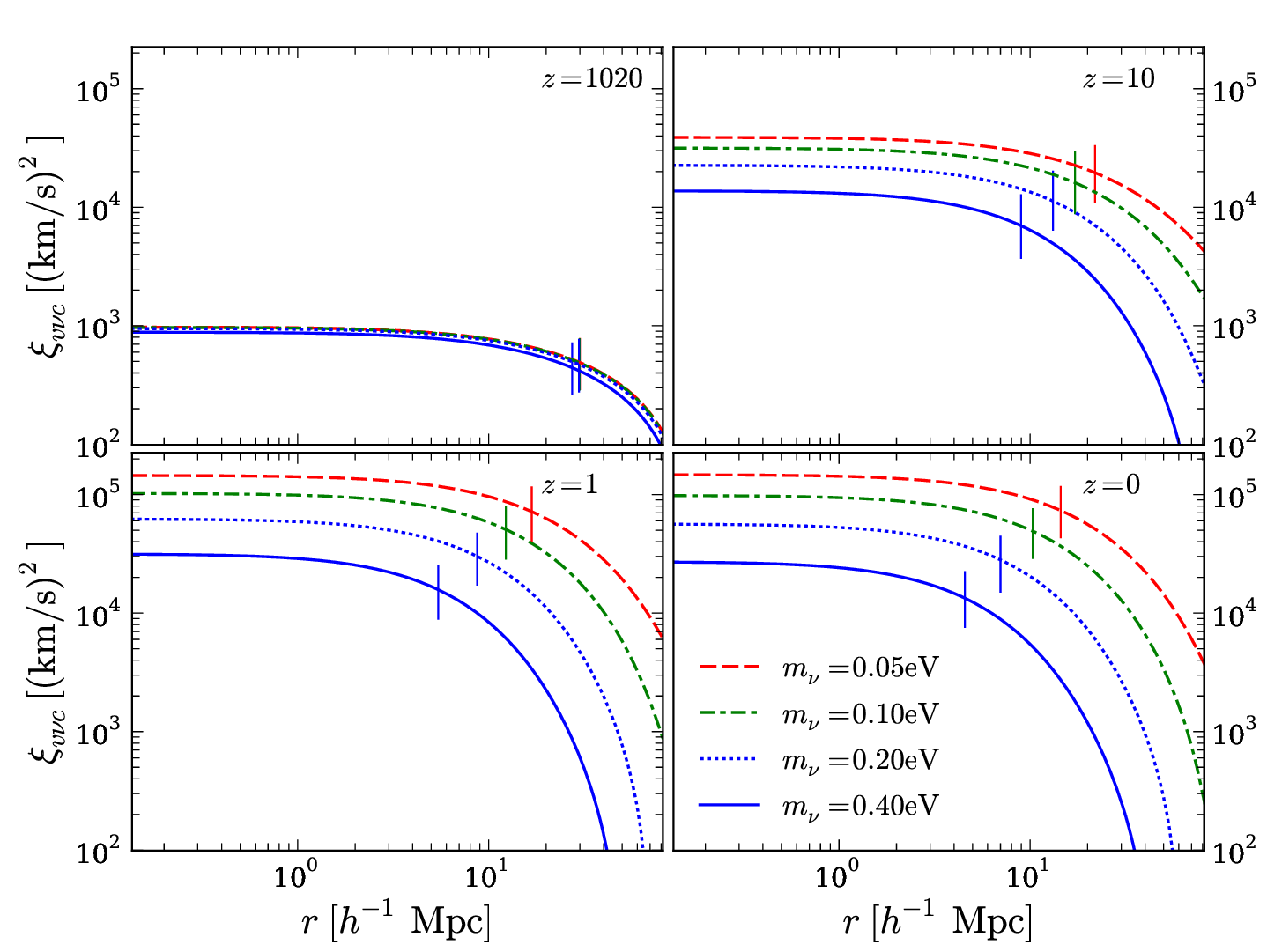}
\end{center}
\vspace{-0.7cm}
\caption{\label{fig:delta_v}
The relative flow correlation function $\xi_{v\nu c}(r)$ at different redshifts. 
The amplitude and scale of the relative flow depends on neutrino mass.
The tick marks the correlation length.}
\end{figure}

The relative velocity correlation function 
$\xi_{v\nu c} (r) 
\equiv \langle \bm{v}_{\nu c}(\bm{x})\bm{v}_{\nu c}(\bm{x}+\bm{r})\rangle$
for four redshifts are shown in Fig.\ref{fig:delta_v}.
 The bulk velocity correlation functions
for different neutrino masses are almost identical at very high redshifts, 
but become increasingly differentiated at low redshifts, as the correlation functions of the lighter neutrinos have larger 
amplitudes and longer correlation 
lengths. The coherent scales $R$, which is defined as the scale at which 
the correlation function $\xi_{v \nu c}$ drops to 
half of its maximum value, are 14.5, 10.3,
7.0, and 4.6 $\Mpc/h$, respectively, for the four neutrino masses at $z=0$.
However, the neutrinos are not visible, so we cannot 
use this correlation function to measure neutrino mass directly. 

{\it Power spectra and correlation functions.}---Because of the bulk relative 
velocity between 
the CDM and neutrinos, the reflection symmetry along the direction of the 
flow is broken locally, and within a velocity coherent 
region the cross-correlation  contains a dipole term, 
\begin{equation}
\xi_{c\nu}(\bm{r}, \bm{v}_{\nu c}^{(bg)})=\xi_{c\nu0}(r, v_{\nu c}^{(bg)})+
\mu\xi_{c\nu1}(r, v_{\nu c}^{(bg)}),
\label{eq:xi_exp}
\end{equation}
 where $\mu=\bm{r}\cdot\bm{v}^{(bg)}_{\nu c}$.
 This also appears as an imaginary part in the CDM-neutrino cross power 
spectrum:
$P_{c\nu}(k, v_{\nu c}^{(bg)}, \mu)= P_{c\nu0}(k, v_{\nu c}^{(bg)}) + i \mu P_{c\nu1}(k, v_{\nu c}^{(bg)})$.
[We can see this by noting that when taking the Hermite conjugate of $P_{c\nu}$,
the imaginary part changes sign and so the angular dependent part
is antisymmetric in ``$c\nu$,'' i.e., $\xi_{\nu c}(\bm{r}, v_{\nu c}^{(bg)})=
\xi_{c\nu0}(r, v_{\nu c}^{(bg)})-\mu\xi_{c\nu1}(r, v_{\nu c}^{(bg)})$.]
This imaginary term would otherwise be zero if not for 
the relative flow between neutrinos and CDM.
This effect is similar to gravitational redshift \cite{2009JCAP...11..026M}, 
which breaks the reflection symmetry along the line of sight, and causes 
an imaginary part in the cross power spectrum
between two types of galaxies.

\begin{figure}[tbp]
\begin{center}
\includegraphics[width=0.48\textwidth]{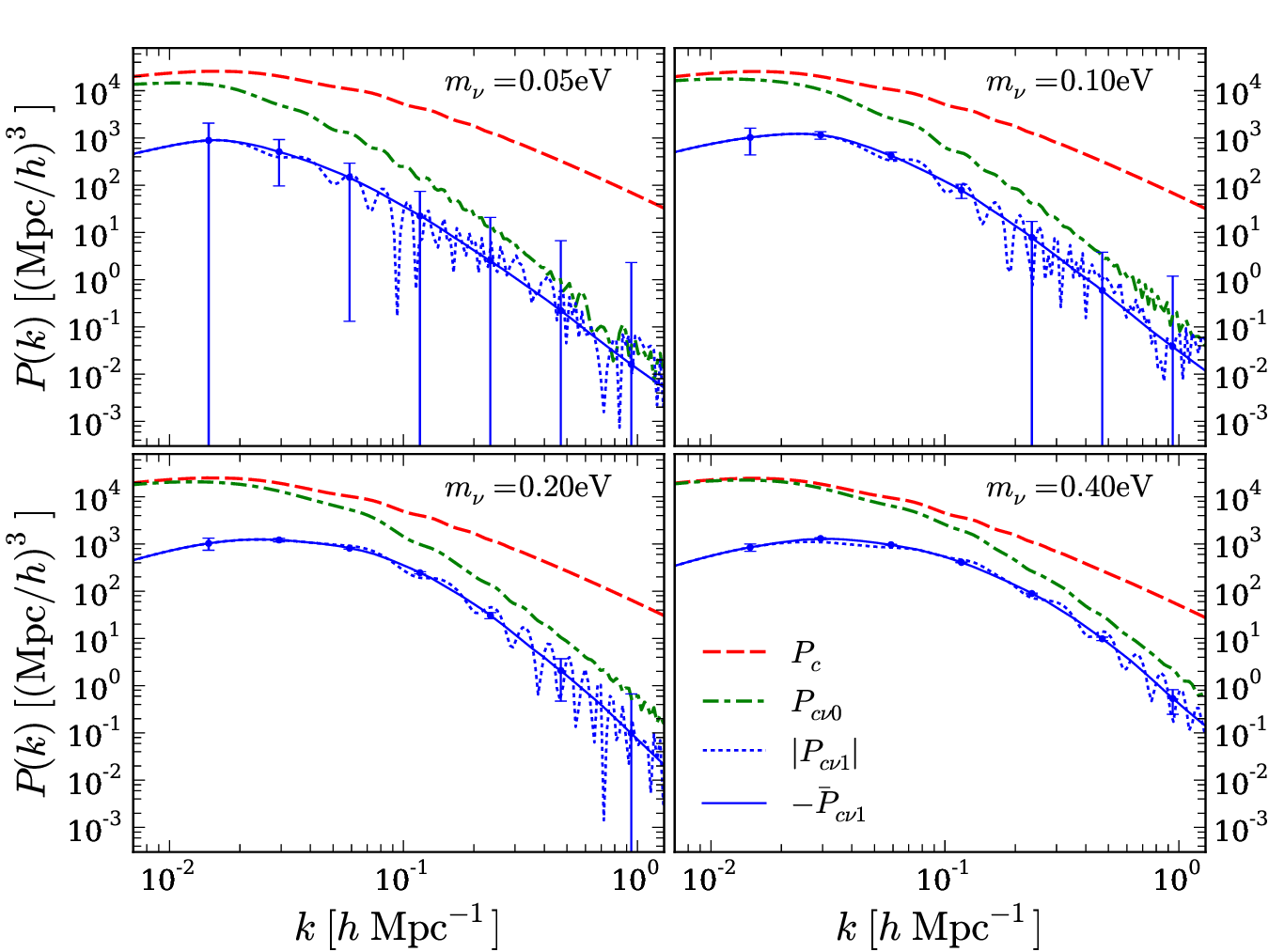}
\end{center}
\vspace{-0.7cm}
\caption{\label{fig:crosspk} The power spectra of
CDM and neutrinos. The CDM auto-power $P_c$, neutrino 
monopole $P_{c \nu 0}$ and dipole $P_{c\nu 1}$, and smoothed 
dipole term $\bar{P}_{c\nu 1}$ are plotted.
}
\end{figure}

\begin{figure}[!htbp]
\begin{center}
\includegraphics[width=0.48\textwidth]{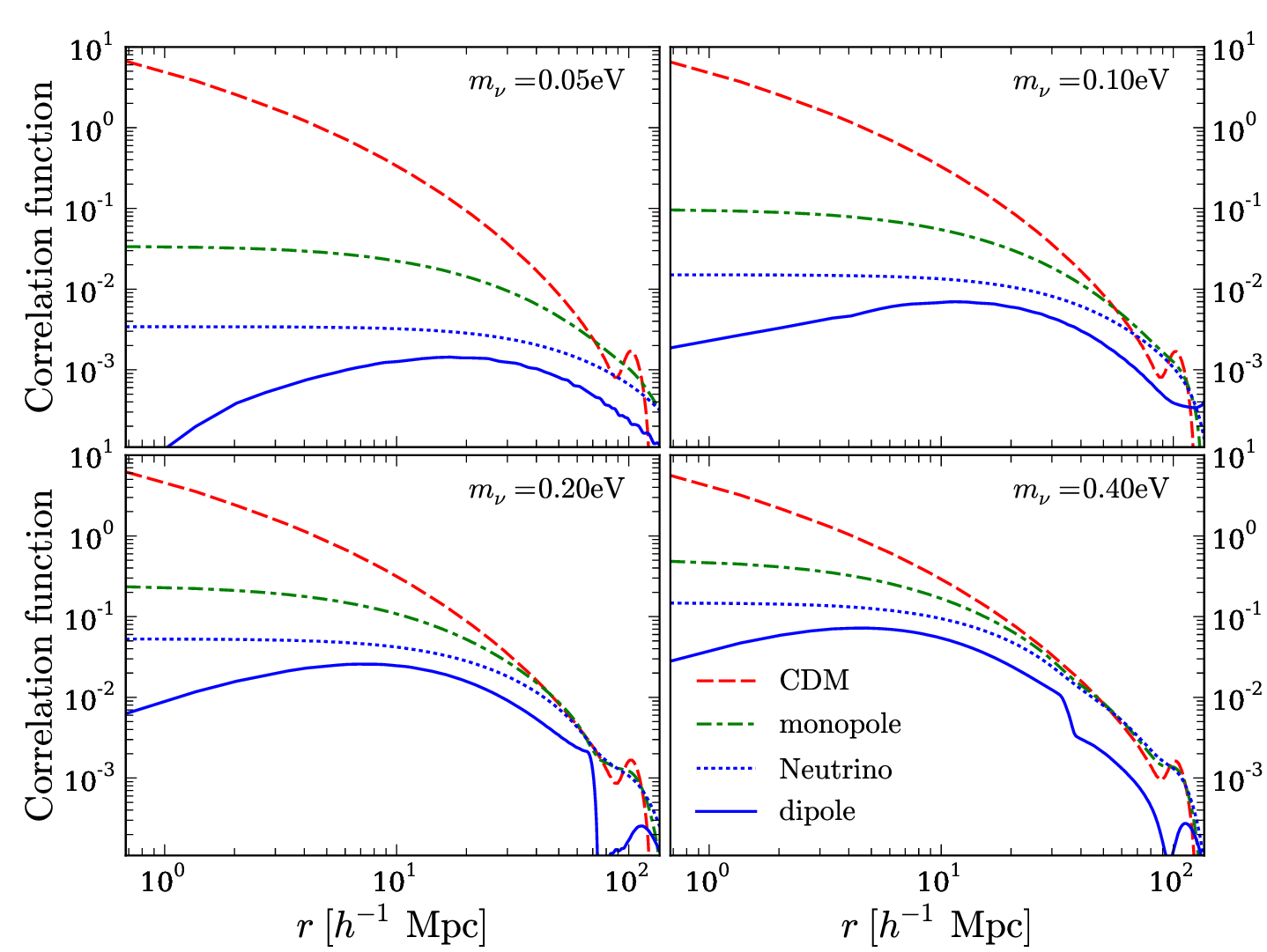}
\end{center}
\vspace{-0.7cm}
\caption{\label{fig:corr}  The correlation functions, including CDM and 
neutrino autocorrelations, and the monopole and dipole part of their 
cross correlations. The dipole is   
$\Delta_{\mathrm{corr}}\equiv\xi_{c\nu}|^{\mu=1}_{\mu=-1}$. }
\end{figure}

Taking $\sqrt{\langle v^2_{\nu c}\rangle}$ as the representative
value for the background velocity, we calculate the induced density 
correlations using MBPT. 
Figure \ref{fig:crosspk} shows the monopole and
the absolute value of the dipole (most parts of it are negative)
terms of the CDM-neutrino cross power spectrum as well as the CDM autopower 
spectrum for four different neutrino masses. 
The oscillations in $P_{c\nu 1}$ (dotted line) 
are due to the sharp sound horizon which is an artifact of the fluid 
approximation of neutrinos in our 
calculation. (We have verified that the oscillation
period is inversely proportionate to the effective sound speed, so it is 
due to the (false) acoustic oscillation in the fluid. Real neutrinos
are not a collisional fluid, and the effective sound speed is actually a
superposition of different sound speeds, so we do not expect the true
cross power spectrum to exhibit these oscillations.) 
We have thus smoothed the dipole
power spectrum and obtained an average $\bar{P}_{c\nu 1}$, which 
is shown as the solid line, for the different neutrino masses the
power spectra are different and distinguishable. 
Figure \ref{fig:corr} shows, respectively, the 
CDM autocorrelation function, the neutrino autocorrelation function,
and the monopole and dipole part of CDM-neutrino cross 
correlation functions. We find that the neutrino autocorrelation 
grows as the neutrino mass increases,
since the more massive neutrinos tend to form more structures. 
The dipole term of the cross power spectrum have a broad peak or hump, 
its amplitude also grows with the neutrino mass.
The scales of the peaks in the correlation function decrease with neutrino
mass, and are located at 16, 11, 7, and 5 Mpc/$h$, respectively, for the
four neutrino masses.  

We have taken a single value of $v^{(bg)}=\sqrt{\langle v^2_{\nu c}\rangle}$ 
for each neutrino mass. For a given background velocity value, 
the dipole correlation depends on the neutrino mass value, as is shown
in the equations in the Supplemental Material \cite{SuppMat}. 
But in fact the bulk relative
velocity varies from point to point in space.
A more rigorous treatment would require a consideration of the distribution
of the bulk velocity.  The fact that both the typical value of 
bulk velocity and the dipole correlation for a given background 
velocity depend on the neutrino mass enhances the sensitivity for this
technique. Below for simplicity we will consider only the typical 
values.

\textit{Observability.}---Neither the neutrinos nor the dark matter can be 
observed directly, but as their densities affect galaxy densities,
their cross power can be inferred from the cross power of galaxies
of different populations, provided that the biases 
of the two populations have different dependences on neutrinos and dark matter.
Galaxies are known to be biased relative to each
other \cite{2009MNRAS.392..682C}.  
The 21cm HIPASS galaxies typically have a
bias of $b_c\sim 0.7$ \cite{2007ApJ...654..702M} relative to the dark matter, 
whereas the bias for luminous red galaxies is typically greater than 1. 
For a galaxy population, we assume its density contrast is related to 
the dark matter and neutrino density contrasts $\delta_c,\delta_\nu$ 
as 
$\delta_g = b_c f_c \delta_c+ b_\nu f_\nu\delta_{\nu}$,
where $f_c = \Omega_c/(\Omega_c+\Omega_\nu)$ and 
$f_\nu = \Omega_\nu/(\Omega_c+\Omega_\nu)$ \cite{2014arXiv1405.4855L}. 
Since the halo mass scale $10^{12}\sim10^{13}M_\odot$  
is smaller than the neutrino free streaming and coherent scales, 
we expect the neutrino bias to be insensitive to halo mass. 
This can also be seen by deriving the halo bias via the peak-background
split formalism or the extended Press-Schechter formalism 
(see, e.g., Ref. \cite{MBWbook}) with neutrino fluctuations only affecting the
large scale background density.
For the following calculations,  we choose $b_\nu$ to be 1 but emphasize 
that an effect will be present as long as $b_\nu$ is the same for both galaxy
populations, regardless of the particular value.
The precise value could be calculated with the more elaborate treatment 
as prescribed in Ref. \cite{2014arXiv1408.1081L}.

If we consider the cross-correlation of 
two galaxy populations denoted by $\alpha, \beta$, and use $b_\alpha,b_\beta$
to denote $b_c$ for $\alpha, \beta$, then
$$\xi_{\alpha\beta}=\langle \delta_\alpha \delta_\beta \rangle = b_\alpha b_\beta f_c^2 \xi_c +
(b_\alpha + b_\beta) f_c f_\nu \xi_{c\nu} + f_\nu^2 \xi_\nu.
$$
Now consider the $\mu$ dependence of $\xi_{\alpha\beta}$:
because the cross correlation function is antisymmetric in ``$c\nu$,''
a dipole $\mu(b_{\alpha}-b_{\beta})f_cf_\nu\xi_{c\nu1}$
appears. The observability of this dipole depends on the
relative bias, $\Delta b\equiv b_{\alpha}-b_{\beta}$.
The known spread in formation
bias provides a lower bound on $\Delta b \gtrsim 0.5$.   For sensitivity
estimation, we will adopt $\Delta b=1$.  The actual error bar of the
inferred neutrino mass will depend on the product of $\Delta b$ and
galaxy number density $n_g$.

For this measurement, the bulk velocity field can be reconstructed
from the observed density field,
\begin{equation}
\bm{v}_{\nu c}(\bm{k})=\delta_g(\bm{
  k}) \frac{[T_{\theta,\nu}(k)-T_{\theta,c}(k)]}{T_{\delta,g}(k)}
  \frac{i\bm{k}}{k^2}.
\label{vnu}
\end{equation}
Here $T_{\theta,\nu}(k), T_{\theta,c}(k)$ are the velocity-divergence transfer functions for
neutrino and dark matter respectively, and $T_{\delta,g}(k)$ is the
density transfer function, which depends on the unknown neutrino mass. In 
practice, one can iterate the reconstruction with different trial masses
$m_\nu$, until a self-consistent relative velocity field
$v_{\nu c}$ and dipole value is found.
At the high sampling densities considered here, the fractional error in
$v_{\nu c}$ is comparable to the error in the CDM density field
$\delta$.  The shot noise is much smaller than the
sample variance, making the error on the velocity field
negligible at the scales of interest.

The correlation function provides a local operational 
procedure to measure the dipole,
\begin{equation}
\xi_{\alpha\beta}(r,\mu)=\frac{1}{N}\sum_{\bm x} \sum_{\begin{subarray}{c}
|\Delta \bm{x}| \sim r \\
\hat{v}_{\nu c}\cdot
  \hat{\Delta x}\sim \mu \end{subarray}}\delta_\alpha(\bm{x})\delta_\beta(\bm{x}+\Delta\bm{x}),
\end{equation}
where $N$ is appropriate normalization.
The dipole term can be extracted from this anisotropic correlation 
as in Eq.(\ref{eq:xi_exp}).
Taking the Fourier transform then yields the power spectrum dipole.  
The error bar is easier to specify for the power spectrum 
than the correlation function, 
since $k$ bins are statistically independent. The
transformation from real space to
redshift space does not change our error estimate because
the dipole is orthogonal to the effect of redshift
distortion, which is a quadrupole distortion.

In  Fig.~\ref{fig:crosspk}, we plot the expected error bars of the
angular-dependent CDM-neutrino cross power spectrum for a survey with
volume $V_s=1.0h^{-3}\mathrm{Gpc}^3$ and $n_g \Delta
b=2.4\times10^{-2}h^3\mathrm{Mpc}^{-3}$. 
This corresponds to an all-sky survey out to
redshift $z<0.2$, comparable to the sloan digital sky survey (SDSS) main 
sample volume, but with
a tenfold higher galaxy sampling density, about the
density of HIPASS galaxies \cite{Zwaan:2003}. The two populations of galaxies
could be, for example, a deep optical survey and
an HI survey at low redshifts. Alternatively, the second tracer might
be obtained by a nonlinear weighting of the same density
field such as the cosmic tide field \cite{2012arXiv1202.5804P}.

We proceed to calculate the error on
the neutrino mass measurement using a Fisher matrix estimate.
We use five $k$ bins ($k=0.059$, 0.12, 0.24, 0.47, 0.94 $h$/Mpc)
in Fig.~\ref{fig:crosspk}. Modes with smaller $k$ are not used
because MBPT is not a very
good approximation unless the background velocity
comes from scales larger than the $k$ mode.  We fit for
two parameters: a multiplicative
(relative) galaxy bias $\Delta b$, treated as a nuisance parameter,
and a neutrino mass, and marginalize the result over the relative bias.
The result is given in Table ~\ref{tab:mnu_error} for
the four different neutrino masses.  Existing galaxy redshift data may
result in a detection for optimistic neutrino mass and bias parameters.  
Future surveys can measure the neutrino masses precisely.

\begin{table}
\caption{\label{tab:mnu_error} The forecasted error on neutrino mass with
a survey of  $V_s=1.0h^{-3}\mathrm{Gpc}^3$, $n_g=2.4\times10^{-2}h^3\mathrm{Mpc}^{-3}$
 and with current survey
data, modeled with SDSS and 2dF as $V_s=0.2h^{-3}\mathrm{Gpc}^3$, $n_g
V_s=1\times10^{6}$. Note that substantial uncertainties exist due to
unknown galaxy neutrino bias, which is a nuisance parameter that we
marginalize over.}
\begin{tabular}{c|cc|cc}
&\multicolumn{2}{c|}{
current (SDSS)}
&\multicolumn{2}{c}{ future} \\
$m_\nu$ (eV) & $\sigma_{m_\nu}$ & relative error & $\sigma_{m_\nu}$ & relative error \\
\hline
0.05& 0.045 & 0.90  & 0.0042 & 0.084\\
0.10& 0.044 & 0.44 & 0.0041 & 0.041\\
0.20 & 0.079 & 0.40& 0.0074 & 0.037\\
0.40 & 0.097 & 0.24& 0.0091 & 0.023\\
\hline
\end{tabular}
\end{table}

\textit{Discussions.}---The neutrino mass measurement method proposed here 
differs from the one based on small scale 
power spectrum suppression, and it is more robust
to scale-dependent galaxy biasing.  In the approach based on power suppression, 
if for some reason there is a weak scale-dependent variation of bias 
at the level of $\sim 1\%$, it can completely swamp 
the neutrino signal. In our dipole
cross correlation approach, the measured signal arises only from
the relative velocity effect.  If the galaxy bias were to
depend on scale, the impact on the inferred neutrino mass would only
be proportionate to any such changes, unlike for total power
measurements where any uncertainty in bias is amplified by 2 orders
of magnitude or more.

For the cases we considered, the correlation function peaks 
occur at scales ($16, 11, 7, 5$ Mpc/h) 
comparable to the relative velocity field coherency 
scales ($14.5, 10.3, 7.0, 4.6$ Mpc/h); this is not unexpected as it is 
the coherence of the bulk velocity which induces such correlation. However, 
for the analytical MBPT calculation we used here, it does pose a problem, 
because strictly speaking the MBPT approximation is valid 
only for scales below the coherence scale.
 The nonlinear effects become significant for $k\gtrsim 0.1 h$/Mpc.
Nevertheless, the essence of large scale velocity modulation 
and the expected physical effect (the dipole structure) is still captured in the
calculation, though quantitatively it may not be very accurate at the 
largest scales. This can be remedied with numerical simulations. We will
study this in a future paper; preliminary results, however, 
show that the result is 
generally consistent with the analytical one.

 In our Fisher analysis, we have
treated the galaxy relative bias as a nuisance parameter.  As described above,
the sensitivity to this effect depends on $n_g\Delta b$ and so the
galaxy density needed to detect this dipole depends on the bias.  In
any given detection of the dipole, $\Delta b$ is immediately known,
and thus the error on the neutrino mass would also be known.  The
uncertainty in the bias, and thus the error, is proportionate to the
significance of the detection, i.e. for a $10\sigma$ detection, there
is an additional $10\%$ uncertainty in the error itself.

In the above we have considered a single neutrino mass. In fact, 
unlike the power spectrum suppression effect, which is sensitive only to the
sum of the neutrino masses, the dipole effect discussed here can in principle
be used to measure the mass of a single neutrino. For multiple neutrinos, 
the different mass eigenstates will have different bulk velocity 
directions for each of them, which at least in theory can be
solved independently by repeating this procedure once for each mass.
In practice this may be difficult, but if one or two neutrino masses 
are dominant and degenerate, then the procedure discussed in this Letter 
is already sufficient. For an inverted neutrino mass hierarchy, 
the effect would be twice as large and enhance the possibility of detection.  

We benefitted from helpful discussions with Joel Meyers and Camille Bonvin.
We acknowledge the support of the Chinese MoST 863
program under Grant \mbox{No. 2012AA121701}, the CAS Science Strategic Priority 
Research Program XDB09000000, the NSFC under Grant \mbox{No. 11373030}, 
Tsinghua University, CHEP at Peking University, and NSERC.

\bibliographystyle{apsrev}
\bibliography{neu}

\end{document}